\documentclass[12pt]{article}
\usepackage{amsfonts,amssymb,amsbsy,latexsym,amsmath}
\begin{document}
\large
\newcommand{\be}{\begin{equation}}
\newcommand{\ee}{\end{equation}}
\newcommand{\ba}{\begin{eqnarray}}
\newcommand{\ea}{\end{eqnarray}}
\newcommand{\dalam}{\raisebox{1mm}{\fbox{}{}}\;}
\newcommand{\pa}{\partial}
\newcommand{\s}{\sqrt}
\let\f\frac
\newcommand{\al}{\alpha}
\newcommand{\st}{\stackrel}
\newcommand{\tk}{\tilde\kappa}
\newcommand{\ep}{\epsilon}
\newcommand{\ds}{\displaystyle}
\newcommand{\ed}{\end{document}}
\newcommand{\ul}{\underline}
\begin{center}
{\bf GRAVITATIONAL WAVES in RELATIVISTIC THEORY of GRAVITATION}\\
\vspace*{0.2cm}
{\it S.S. Gershtein, A.A.Logunov and M.A. Mestvirishvili}

\vspace*{0.6cm}

{Abstract}
\end{center}

\vspace*{0.2cm}

{\small It is shown that gravitational waves do not have nonphysical
``ghost'' states in the
Relativistic Theory of Gravitation with graviton having nonzero rest
mass
due to the causality condition.}

\vspace*{0.6cm} It was shown in [1,2] that, in linearized  theory
of gravitation, introducing the rest mass of  graviton for a field
with spins 2 and 0 leads to nonphysical ``ghost states'' due to
spin 0 when interpreting gravitational effects in the Solar
system. ``Ghost'' states appear also in the gravitational
radiation. A conviction has grown up on this basis that just this
proves that the graviton mass is exactly zero. However, in their
study the authors of [1,2] do not treat the gravitational field as
a tensor \textbf{physical field} in Minkowski space which
generates an effective Riemannian space, and as a consequence, its
own causality cone. That is why the \textbf{causality condition}
has not appeared whereas it arises in the Relativistic Theory of
Gravitation (RTG)~[3], because it has two causality cones. There
is only one causality cone of the Riemannian space in General
Relativity.

In the study of gravitational radiation it was found in paper [4]
that taking into account nonlinear terms allows one to eliminate
``ghost'' states. This treatment was admitted also by us and it
was given in [3]. In this work we have obtained that, in RTG, the
negative energy flow and, hence, ``ghost'' states are eliminated
even without an account of nonlinear terms. It turns out to
be<неиsufficient to fulfill the causality condition from the RTG.
Following papers [5,6,7] We start from the opportunity of
existence of a free gravitational field -- the gravitational
waves, as an objective physical reality similar to electromagnetic
waves in vacuum.

For the simplicity and accuracy of our analysis we consider a weak
plane gravitational wave in vacuum with amplitude $a^{\mu\nu}(k)$,
propagating along $Z$ axis \be \Phi^{\mu\nu}= a^{\mu\nu} (k) \cos
kx, \ee where $k_{\nu}=(\omega,0,0,- q\omega)$, $q^2=1-\frac
{m^2}{\omega^2}$, and $m$ is the graviton mass.

We use system of units conventions $G=\hbar=c=1$. In vacuum the
basic RTG equations in linear approximation and in an inertial
frame with Galilean coordinates are taking the following form \be
\mathop{\Box}\Phi^{\mu\nu} + m^2\Phi^{\mu\nu}=0,\ee\be\partial_\nu
\Phi^{\mu\nu}=0. \ee The wave  (1) is a solution of these
equations. A weak gravitational field $\Phi^{\mu\nu}$ produces an
effective Riemannian space with the following metric tensor
$$g_{\mu\nu}=\gamma_{\mu\nu} -\Phi_{\mu\nu} +\frac {1}{2}\gamma_{\mu
\nu}\Phi,\;\;\Phi_{\mu\nu}\gamma^{\mu\nu}=\Phi;$$
tensor $g^{\mu\nu}$ is given by the analogous expression \be
g^{\mu\nu} =\gamma^{\mu\nu}+ \Phi^{\mu\nu} -\frac{1}{2}\gamma^{\mu\nu}
\Phi.\ee
It follows from the above that scalar curvature of the effective
Riemannian space $R$ is $$R=\frac{1}{2}m^2\Phi.$$
But it occurs so that it does not influence the energy flow, as we
shall see below.
Minkowski space interval in an inertial frame with Galilean
coordinates is
\be
d\sigma^2 =\gamma_{\mu\nu} dx^\mu dx^\nu=dt^2-dx^2-dy^2-dz^2.\ee
As RTG treats the gravitational field as a physical tensor field
propagating in Minkowski space,
the causality cone of the arising effective Riemannian space should
not go out the causality cone of the
Minkowski space.  Just this is  the causality principle of the RTG.
According to this principle, the
timelike and lightlike geodesics of the effective Riemannian space
which is produced by the physical
field should not go outside boundaries of the Minkowski space cone.
Just this physical requirement
should get the proper mathematical   formulation.

The terms with second derivatives over spacetime coordinates
appear in the  {\bf hyperbolic dynamical equations} of the
gravitational field in RTG in the following form\be g^{\mu\nu}\;
\frac{\pa^2\Phi^{\alpha\beta}}{\pa x^\mu \pa x^\nu}\;.\ee The
characteristic equation for the gravitational equations is
provided by higher order derivative terms (6) only\be g^{\mu\nu}\;
\frac{\pa S}{\pa x^\mu}\; \frac{\pa S}{\pa x^\nu}=0.\ee This
equation determines  wavefront  of  the field, if graviton have no
rest mass. {\bf The characteristics determine the causality cone
of effective Riemannian space.} Each term with a second derivative
from (6) has the corresponding term in characteristics  (7). If
some term with a second derivative is absent in (6), then there
will be no corresponding term in (7).

The timelike geodesic lines in correspondence with (7) are given
by the Hamilton-Jacobi equations\be g^{\mu\nu} \frac{\pa S}{\pa
x^\mu}\;  \frac{\pa S}{\pa x^\nu}=1.\ee The total set of geodesic
lines in correspondence with (6) is determined by the following
equations
$$
g^{\mu\nu} \frac{\pa S}{\pa x^\mu}\; \frac{\pa S}{\pa x^\nu}=\left \{
\begin{array}{r}0\\1\\-1\end{array}\;,\right.
$$
where the first equation gives the isotropic geodesics, the second
-- timelike geodesics, whereas the third gives spacelike geodesic
lines.

Therefore, on the base of Eqs. (7) and (8) isotropic and timelike
geodesic lines, in correspondence with Eq. (6), fulfill the
following inequality \be g^{\mu\nu} \frac{\pa S}{\pa x^\mu}\
\frac{\pa S}{\pa x^\nu}\geq 0.\ee This inequality may be written
as follows\be g_{\al\beta} p^{\alpha}p^\beta \geq 0,\ee where
contravariant vector $p^\al$ is\be p^\al =g^{\al \mu} \frac{\pa
S}{\pa x^\mu}.\ee To provide that the causality cone of the
effective Riemannian space  be inside the causality cone of
Minkowski space it is necessary and sufficient to fulfill the
following inequality \be\gamma_{\al\beta} p^\al p^\beta \geq 0.\ee
Inequalities (10) and (12) may be written in a form directly
connected with the geodesic motions (7) and (8)  which are in
exact correspondence with  (6)\be g^{\mu\nu} p_\mu p_\nu \geq
0,\ee\be\gamma_{\al\beta} g^{\al\mu} g^{\beta\nu}p_\mu p_\nu \geq
0,\ee where covariant vector $p_\nu$ is\be p_\nu=\frac{\pa S}{\pa
x^\nu}.\ee Causality conditions (13) and (14) put definite rigid
restrictions on solutions of the gravitational field equations.
Only the solutions satisfying inequalities (13) and (14) have a
physical meaning in the theory. Inequalities (13) and (14) {\bf
are straightforwardly connected with the hyperbolic equations for
the gravitational field} as they are derived from {\bf the second
derivatives structure (6)} of the dynamical equations. Just this
mathematical formulation of the causality principle guarantees the
position of the Riemannian causality cone inside the causality
cone of the Minkowski space, in correspondence with the dynamical
structure (6).

Earlier in [3] we have not recognized this fact of necessity to
establish the direct correspondence of the causality principle
with the {\bf hyperbolic dynamical system of equations} . In the
case of static system the gravitational field equations are not
hyperbolic and so such a direct correspondence is absent. But in
that case the causality condition can be used in the form of
inequalities (10) and (12). Taking into account
$$
\tilde g^{\mu\nu}=\tilde \gamma^{\mu\nu} +\tilde\Phi^{\mu\nu},
$$
where
$$\tilde g^{\mu\nu} =\sqrt{-g} g^{\mu\nu},\;\;\tilde \gamma^{\mu\nu} =
\sqrt{-\gamma}
\gamma^{\mu\nu},\;\;\tilde \Phi^{\mu\nu} =\sqrt{-\gamma}
\Phi^{\mu\nu},$$
inequalities (13) and (14) in inertial frame with Galilean coordinates
take the following form
\be(\gamma^{\mu\nu} + \Phi^{\mu\nu}) p_\mu p_\nu \geq 0,\ee\be\gamma_{
\al\beta} (\gamma^{\al\mu} +
\Phi^{\al\mu})\;(\gamma^{\beta\nu} +\Phi^{\beta\nu})p_\mu p_\nu \geq
0.\ee
For the motion (1) the following characteristic equation is valid
$$g^{\mu\nu}p_\mu p_\nu = g^{00} (p_0)^2+2g^{03}p_0p_3+g^{33} (p_3)^2=
0.$$
For the weak gravitational field and {\bf the special motion (1) along
$Z$-axis} inequality (16) is fulfilled if
the value of $x$ defined as$$x=\frac{p_3}{p_0},$$is limited by the
following inequalities
\be
x_1\leq x\leq x_2,\ee
where
\begin{eqnarray}&&   x_1 =\Phi^{03} -1 -\frac{1}{2}(\Phi^{00} +\Phi^{
33}),
\nonumber \\[-0.55mm]\\[-0.55mm]&& x_2=\Phi^{03}+1 +\frac{1}{2}(\Phi^{
00}+\Phi^{33}).
\nonumber\end{eqnarray} So we have defined the set of timelike
vectors laying inside the causality cone determined by
characteristics on the base of (6) for the motion (1), leading to
metric (4). Inequality (17) will be fulfilled if $$x'_1 \leq x
\leq x'_2,$$
where
\begin{eqnarray}&& x'_1=2\Phi^{03} -1 -\Phi^{00} -\Phi^{33}, \nonumber
\\
[-0.55mm]\\[-0.55mm]&&x'_2=2\Phi^{03}+1+\Phi^{00}+\Phi^{33}. \nonumber
\end{eqnarray}
In order to provide the position of the effective Riemannian
causality cone inside the Minkowski causality cone it is necessary
and sufficient to fulfill the following inequalities\be x'_1\leq
x_1,\;\; x_2\leq x'_2.\ee On the base of (21) and taking into
account (19) and (20) we obtain \be\Phi^{00} \pm 2\Phi^{03}
+\Phi^{33} \geq 0.\ee From Eq. (3) we find for the solution (1) :
\be\Phi^{00}=q\Phi^{03}, \;\; \Phi^{03}=q\Phi^{33}.\ee After
substituting these equations into (22) we obtain\be(q\pm
1)^2\Phi^{03} \geq 0.\ee It follows from these inequalities for
the wave (1) that \be\Phi^{03}\equiv 0,\ee and therefore, on the
base of Eqs. (23), the following equations take place
\be\Phi^{00}\equiv 0,\;\; \Phi^{33}\equiv 0.\ee In RTG the
causality principle selects the physical solution of the
gravitational equations. It follows from (25) and (26) that
longitudinal-longitudinal components are absent in the wave
solution (1). Just for this reason there are no term
like$$R\Phi^{03} =\frac{1}{2} m^2 \Phi\Phi^{03},$$in the energy
flow, this term is identically zero.When the graviton mass is
zero, Eqs. (25),(26)  as a rule are derived from the gauge
transformations. Here they follow from the causality principle.
Just this provides the positivity of the energy flow in RTG in
case of the nonzero graviton mass.

In the RTG quadratic approximation considered in Galilean
coordinates the energy flow is determined, according to [3,4], by
means of the \textbf{tensor} quantity \be t_g^{\epsilon\lambda}
=\frac{1}{32\pi}\gamma^{\epsilon\alpha}\gamma^{\lambda\beta}\left
(\partial_\alpha \Phi^\tau_\nu\cdot\partial_\beta
\Phi_\tau^\nu-\frac{1}{2}\partial_\alpha\Phi\cdot\partial_\beta
\Phi\right ). \ee Rising and lowering of the indices for
$\Phi^{\mu\nu}$ is provided by means of metric tensor
$\gamma_{\mu\nu}$. According to Eq. (3), the following relations
take place for solution (1):
\begin{eqnarray}&& a^{10}=qa^{13} \nonumber \\[-0.55mm]\\[-0.55mm]&& a
^{20}=qa^{23}. \nonumber\end{eqnarray}
Taking into account (1), and also Eqs. (25), (26) and (28), on the
base of Eq. (27) we obtain for the wave (1) after averaging over
time\be t^{03}_g=\frac{1}{32\pi}q\omega^2\left \{(a^2_1)^2
+\frac{1}{4}(a^1_1-a^2_2)^2+ \frac{m^2}{\omega^2}\left
[(a^1_3)^2+(a^2_3)^2\right ]\right \}.\ee It follows from here
that only transverse-transverse components  are present in the
density of the flow for the wave (1), and longitudinal-transverse
$a^1_3$, $a^2_3$, $a^1_0$, $a^2_0$ also. The last ones are
multiplied by $\frac{m^2}{\omega^2}$ in the energy flow (29).The
longitudinal-longitudinal components {\bf are absent} in the wave
(1). It should be noted that according to RTG it is possible to
provide a continuous transformation to the zero graviton mass in
this problem.

Therefore it follows from (29) that the presence of nonzero
graviton mass does not lead in the RTG to the appearance of
nonphysical ``ghost'' states. The ``ghost'' states also do not
appear in the RTG when explaining the Solar system effects. Here
there is a continuous transformation to the zero graviton mass at
the distance from the source $r\gg r_g=2M$. The graviton mass
arises in the RTG with the necessity when we begin to treat the
gravitational field as a physical one in Minkowski space.

The authors are grateful to V.I.~Denisov, V.A.~Petrov,
A.P.~Sa\-mokh\-in, K.A.~Sveshnikov, N.E.~Tyurin for valuable
discussions.

\vspace*{0.2mm}

\begin{center}

REFERENCES

\end{center}

\noindent1. Zakharov V.I. JETP Letters, Vol. 12:9 (1970), pp.
312-314.

\noindent2. H. van Dam and M. Veltman. Nuclear Physics B22 (1970),
pp. 397-411.

\noindent3. Logunov A.A. Relativistic Theory of Gravitation.
Moscow, Nauka Publishers, 2006 (in Russian).

\noindent4. Loskutov Yu.M. Theor. Math. Phys. Vol.107:2 (1996),
pp. 686-697.

\noindent5. Landau L.D. and Lifshits E.M. Field Theory. Moscow,
Fizmatlit Publishers, 2001 (in Russian).

\noindent6. Eddigton A.S. Theory of Relativity. Moscow-Leningrad,
1934 (in Russian).

\noindent7. Einstein A. Collection of Works. Vol. 1, Moscow, Nauka
Publishers, 1965 pp. 631-646 (in Russian). [Uber
Gravitationwellen, Sitzungsber. preuss. Akad. Wiss., 1918, 1,
154-167]\enddocument